\documentclass[aps,prl,twocolumn,superscriptaddress]{revtex4-2}
\usepackage{amsmath}
\usepackage{graphicx}
\usepackage{mathtools}
\usepackage{mathtools}
\usepackage{float}
\usepackage{empheq}
\usepackage{epstopdf}
\usepackage{bigints}
\usepackage{epstopdf}
\usepackage{amssymb}
\usepackage{wrapfig}
\usepackage{amsmath,hyperref}
\usepackage{cleveref}
\usepackage{soul}
\usepackage{textcomp}
\usepackage{mydef}
\usepackage{xcolor}
\usepackage[makeroom]{cancel}
\usepackage{float,graphicx}
\hypersetup{
colorlinks,
citecolor=blue,
filecolor=red,
linkcolor=blue,
urlcolor=blue,
hyperfigures
}
\begin{document}
\title{Hydrodynamic instabilities in driven chiral suspensions}
\author{Seema Chahal}
\author{Brato Chakrabarti}
\affiliation{International Centre for Theoretical Sciences, Tata Institute of Fundamental Research, Bengaluru, 560089, India} 
\email{seema.s@icts.res.in}
\email{brato.chakrabarti@icts.res.in}

\date{\today}
	
\begin{abstract}	
Active Stokesian suspensions are conventionally understood to generate dipolar stresses that destabilize aligned states in the bulk and drive system-wide spatiotemporally chaotic flows. Here, we report dynamics in suspensions of torque-driven spinning chiral particles that exhibit a distinct and previously unrecognized route to collective dynamics. Using a mean-field kinetic theory, stability analysis, and nonlinear simulations, we demonstrate how flows driven by torque monopoles and self-propulsion resulting from microscopic chirality drive chaotic flows in three dimensions. Unlike the well-known alignment instability of dipolar active matter, the present dynamics is intrinsically tied to self-propulsion and relies on the emergent coupling between nematic and polar order. Our results establish a novel route to pattern formation, suggest strategies for designing torque-driven active suspensions, and provide a mechanistic framework to probe the rheology of chiral fluids.
\end{abstract}
\pacs{...}
\maketitle

Microorganisms, active cytoskeletal structures, and their synthetic analogs perform mechanical work on their surroundings by consuming chemical energy. This activity typically results in dipolar stresses -- a hallmark of force-free active particles. It is well-known that bulk suspensions of such dipolar active particles are susceptible to the `generic' alignment instability which renders a Stokesian flock unstable beyond a critical activity, leading to states of spontaneous flow and chaotic dynamics \cite{aditi2002hydrodynamic,saintillan2008instabilities,subramanian2009critical,baskaran2009statistical,marchetti2013hydrodynamics,alert2020universal}. This instability has become the hallmark and defining route to collective dynamics in active Stokesian fluids \cite{ramaswamy2010mechanics,saintillan2014theory,alert2022active}. In this letter, we focus on a distinct class of active, or rather driven, suspensions comprising particles that do not produce dipolar stresses. Instead, each particle spins around its body axis under the action of external torque and thus acts as a torque monopole in the bulk fluid \cite{de2024pattern}. For such systems, the generic route to pattern formation is absent. However, we uncover a new pathway to collective dynamics that arises when the particles are chiral. The microscopic screw-like chirality results in self-propulsion along the spinning axis, which in turn triggers an instability unique to self-propelled particles, leading to emergent system-wide flows in three-dimensional suspensions. Our results reveal a previously unrecognized mode of self-organized flows, with potential implications for engineered driven suspensions \cite{ghosh2009controlled,goyal2025externally} and rheology of chiral fluids \cite{banerjee2017odd}.

Chirality is a pervasive feature of biological and synthetic active systems \cite{maitra2025activity}. Examples range from the clockwise swimming of surface-bound \textit{E. coli} \cite{darnton2004moving}, to rotating cilia \cite{smith2008fluid}, and chiral actomyosin flows \cite{furthauer2013active,naganathan2014active} at the cell cortex—each illustrating how molecular chirality shapes emergent large-scale dynamics. Another prominent manifestation of chirality is the coupling between translation and rotation, seen in the spinning propulsion of cytoskeletal filaments in gliding assays \cite{meissner2024helical}, the swimming of \textit{E. coli} driven by their helical flagella \cite{lauga2020fluid}, and the engineered motion of driven helical micromotors \cite{goyal2025externally}. Most prior theoretical and experimental work on collective manifestation of chirality have focused on quasi-two-dimensional settings \cite{soni2019odd}, including layered \cite{maitra2020chiral,kole2021layered} and nematic materials \cite{maitra2019spontaneous}, or spinning collectives confined to a plane \cite{patra2022collective,digregorio2025phase}; only recently has the role of chirality been probed in three-dimensional dry active matter \cite{kuroda2025singular}. While a recent hydrodynamic framework addressed the surface motility of the rod-shaped bacterium \textit{Myxococcus xanthus}, which spins its body to glide on substrates \cite{banerjee2024active}, the dynamics of momentum-conserving bulk suspensions of torqued chiral particles, where rotation and translation are aligned, remain unknown.

\begin{figure*}
\centering 
\includegraphics[width = 1\linewidth]{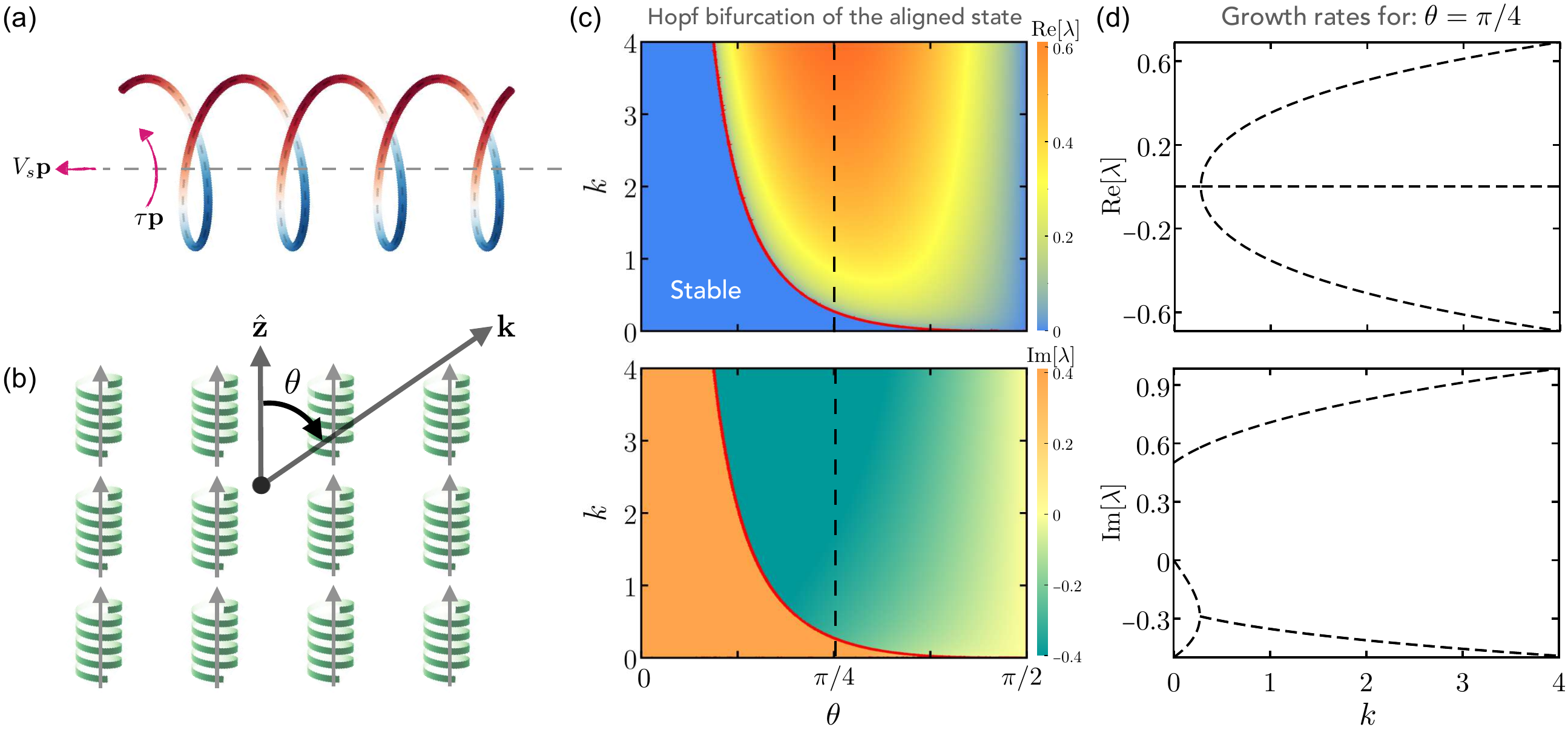}
\caption{\textbf{(a)} Screw-like chiral particles are actuated along their long axis by external torque $\tau \bp$. Translation–rotation coupling causes each particle to both spin and self-propel along $\bp$. \textbf{(b)} Illustration of the uniaxial (polar) phase of the suspension and the perturbation wavevector $\bk$. Because the particle stress is polar, the stability of apolar-nematic states differs from that of the polar states analyzed here (see SI). \textbf{(c)} Dispersion relation of the most unstable eigenvalue in the $k-\theta$ space. The real part of the growth rate (top panel) shows a finite $k$ instability beyond the red stability boundary. Imaginary parts (bottom panel) illustrate that the instability carries the signature of a Hopf bifurcation. \textbf{(d)} Dispersion relation as a function of wavenumber $k$ for a given $\theta = \pi/4$ (dashed line in panel (c)). The three branches correspond to the numerically obtained eigenvalues of the linearized problem. For all the analysis shown here we chose: $d_T = d_r = 0$ and $\chi = 1$.}
\label{fig:Fig1}
\end{figure*}

Here, we combine theory and simulations to unravel a range of new hydrodynamic instabilities for spinning, chiral elongated particles that exert a torque monopole along their long axis (Fig.~\ref{fig:Fig1}\textbf{(a)}). Such a hydrodynamic signature is in contrast with truly active particles (ex., a bacterium) with an internal source of energy that are torque-free and hence exert a torque-dipole in the bulk \cite{lauga2020fluid}. Thus, the particles under consideration in this letter are driven or actuated by external means \cite{goyal2025externally}. One approach to experimentally realizing a torque-driven suspension is to harness electrohydrodynamic instabilities such as Quincke rotations of dielectric particles in a bulk fluid \cite{vlahovska2019electrohydrodynamics}. Such dynamics have been reported experimentally \cite{brosseau2017electrohydrodynamic,lefranc2025quorum} for isolated spheroids and theoretically for chiral particles like helices that propel along their spinning axis \cite{das2019active}. Here, we demonstrate how such microscopic chirality and fluid-mediated long-range interactions resulting from torque monopoles conspire to drive instabilities in dilute suspensions of these chiral particles.

\vspace*{1mm}
 
\noindent \textit{Modeling:} We start by laying out our model, which builds on a mean-field kinetic theory for spheroidal particles suspended in a Stokesian fluid \cite{saintillan2008instabilities}. The suspension is described by a probability density function $\Psi(\bx,\bp,t)$ that encodes the probability of finding a particle at position $\bx$ with director $\bp \in \mathbb{S}^2$ (the unit sphere) characterizing the orientation of the long-axis of the particle at time $t$. The distribution function obeys a conservation equation
\begin{equation}\label{eq:smol_eq}
     \frac{\partial \Psi}{\partial t} + \mathbf{\nabla_x} \cdot (\dot{\mathbf{x}} \Psi) + \mathbf{\nabla_p} \cdot (\dot{\mathbf{p}} \Psi) = 0, 
 \end{equation}
 where $\Psi(\bx,\bp,t)$ is normalized as 
 \begin{equation}
      \frac{1}{V} \int_V \md \bx \int_{\mathbb{S}^2} \md \bp \ \Psi(\bx,\bp,t) = \bar{n},
 \end{equation}
 with $\bar{n}$ being the mean number density of the particles. In Eq.~\eqref{eq:smol_eq}, $\nabla_\bp$ is the gradient operator on the unit-sphere, and $\dot{\bx}$ and $\dot{\bp}$ are respectively the translational and rotational flux velocities given by
 \begin{align}
    \dot{\mathbf{x}} &= V_s \mathbf{p} + \mathbf{u}(\bx) - d_T \mathbf{\nabla_x} \ln \Psi, \label{eq:xdot} \\ 
    \dot{\mathbf{p}}  &= (\bI - \bp \bp) \cdot \left[(\gamma \mathbf{E} + \mathbf{W}) \cdot \bp  - d_r  \nabla_\bp \ln \Psi \right]. \label{eq:pdot}
 \end{align}
In Eq.~\eqref{eq:xdot} $V_s$ is the self-propulsion speed of the particles, $\bu(\bx)$ is the self-generated mean-field velocity arising due to the particle stress in the suspension, and $d_T$ is the translational diffusion coefficient. Note that $V_s \neq 0$ only for chiral particles: due to microscopic chirality, the self-propulsion speed is intrinsically linked and proportional to the actuating torque on the particle. Such a propulsion mechanism differs from the well-studied dipolar active particle, such as a microswimmer, where the swimming speed can be independent of its far-field signature of dipolar flows. To model the orientational dynamics of the particles in Eq.~\eqref{eq:pdot} we used Jeffery's equation \cite{d02f014e-d974-38f6-8a6b-030f72ab06a0}, where $\gamma$ is a geometric parameter characterizing the shape anisotropy and $d_r$ is the rotational diffusion coefficient; $\{\mathbf{E},\mathbf{W}\}$ are the symmetric and antisymmetric part of the mean-field velocity gradient tensor. While Jeffery’s equation is appropriate for homo-chiral particles, recent work has identified higher-order corrections for other chiral shapes \cite{Dalwadi_Moreau_Gaffney_Ishimoto_Walker_2024, Dalwadi_Moreau_Gaffney_Walker_Ishimoto_2024}; chirality may also induce a weak drift of the particle's center-of-mass \cite{marcos2009separation}. We ignore such effects for simplicity. Here, we restrict ourselves to a dilute suspension with $\bar{n} \ell_p^3 \ll 1$, where $\ell_p$ is the characteristic length of our active particles; in this dilute limit, the mean-field flow $\bu(\bx)$ is self-consistently computed from a forced incompressible Stokes equation as
 \begin{equation}\label{eq:stokes}
     -\nabla q + \mu \Delta \bu + \nabla \cdot \bSigma^\text{a} = 0, \ \ \nabla \cdot \bu = 0,
 \end{equation}
 where $q$ is the pressure and $\bSigma^\text{a}$ is a particle stress on the fluid due to the presence of torque monopoles along the axis of our particles. Following \cite{batchelor1970stress}, the volume-averaged particle stress at $\bx$ is given as
 \begin{equation}\label{eq:stress}
     \Sigma_{ij}^\text{a}(\bx,t) = \frac{\tau}{2}\int_{\bS^2} \epsilon_{ijk} p_k \Psi(\bx,\bp,t) \ \md \bp = \frac{\tau}{2} \epsilon_{ijk} 
     c n_k.
 \end{equation}
 Here $\tau$ is the magnitude of the external actuating torque and $\epsilon$ is the Levi-Civita tensor. In Eq.~\eqref{eq:stress}, we have also defined the mean-field density and polarity field as $c = \int_{\mathbb{S}^2} \Psi(\bx,\bp,t) \md \bp$ and $\bn = 1/c \int_{\mathbb{S}^2} \bp \Psi(\bx,\bp,t) \md \bp$. We highlight that each spinning particle in our problem acts as a rotlet and the resulting particle stress $\bSigma^\text{a}$ is anti-symmetric as required from the conservation of angular momentum. Importantly, the particle stress is also polar in contrast to the familiar symmetric nematic active stresses that depend on the nematic order.
 The above equations can be made dimensionless using the following velocity, length, and time scales: $u_c \sim \tau/2\mu \ell_p^2$, $l_c \sim (\bar{n} \ell_p^2)^{-1}$, and $t_c \sim l_c/u_c$. After the above scaling, we are left with dimensionless diffusivities $d_T$ and $d_r$, and a geometric factor $\chi = V_s/u_c \sim \mathcal{O}(1)$ that depends on the microscopic chirality and associated mobility of individual particles, capturing the intrinsic coupling between self-propulsion and the torque monopole (see SI). The number density $\bar{n}$ only appears in our problem through the system size scaled by the $(\bar{n} \ell_p^2)^{-1}$. Equations~\eqref{eq:smol_eq}-\eqref{eq:stress} thus provide us with a self-consistent evolution problem for $\Psi(\bx,\bp,t)$; in this micro-macro framework, the actuation of the particles by external torque forces the Stokesian fluid, which in turn affects their collective distribution. We now examine how these hydrodynamic interactions, combined with microscopic chirality, can drive instabilities and pattern formation in such suspensions.

\vspace*{1mm}

\noindent {\textit{Linear Stability Analysis:}}
To highlight the possible emergent dynamics, we first study the dynamics of a uniaxial, aligned state of spinning particles. In particular, we consider a nematic and polar state of the suspension described by $\Psi(\bx,\bp,t) = c(\bx,t) \delta[\bp - \bn(\bx,t)]$ around which the evolution of $c$ and $\bn$ follows
 \begin{align}
    \frac{\mathrm{D} c}{\mathrm{D}t} + \chi \nabla \cdot ( \bn c ) - d_T \Delta c &= 0,  \label{eq:c_align} \\
    \frac{\mathrm{D} \bn}{\mathrm{D}t} + \chi \bn \cdot \nabla \bn  - d_T \Delta \bn  &= \mathcal{P}_{\bn} \cdot (\gamma \mathbf{E}+\mathbf{W}) \cdot \bn . \label{eq:n_align} 
 \end{align}
Here $\mathrm{D}/\mathrm{D}t = \partial_t + \bu \cdot \nabla$ is the material derivative,  $\mathcal{P}_{\bn}=(\bI - \bn \bn)$ is a projection operator, and we have ignored rotational diffusion. The homogeneous, aligned state with $c = 1$ and $\bn = \bn_0$ is a fixed point of the problem with the mean-field velocity $\bu(\bx) = \mathbf{0}$. To probe the stability of this fixed point we use plane wave perturbations of the following form: $c=1 + \varepsilon \Tilde{c}_k e^{\mathrm{i} \bk \cdot \bx + \sigma t}$ and $\bn= \mathbf{\hat{z}} + \varepsilon \Tilde{\bn}_k e^{\mathrm{i} \bk \cdot \bx + \sigma t}$ with $|\varepsilon| \ll 1$. Here, $\bk$ is the wave vector and $\sigma$ is the growth rate of the perturbation; furthermore, we have assumed $\bn_0 = \uvc{z}$ without any loss of generality with $\tilde{\bn} = \{\tilde{n}_1,\tilde{n}_2,0\}$. Retaining terms up to $\mathcal{O}(\varepsilon)$, we linearize Eq.~\eqref{eq:c_align}-\eqref{eq:n_align} and solve for the perturbed mean-field velocity from Eq.~\eqref{eq:stokes} in Fourier space to obtain an eigenvalue problem for the growth rate $\sigma$ as
\begin{gather}
    \begin{split}
        & \lambda \Tilde{c} = -\mathrm{i} \chi k \cos\phi \sin\theta \Tilde{n}_1 - \mathrm{i} \chi k \sin\phi \sin\theta \Tilde{n}_2, \\&
        \lambda \Tilde{n}_1 = -\sin\phi \cos\theta \sin\theta \Tilde{c} + \cos^2\theta \Tilde{n}_2, \\&
        \lambda \Tilde{n}_2 = \cos\phi \cos\theta \sin\theta \Tilde{c} - \cos^2\theta \Tilde{n}_1.
    \end{split}
\end{gather}
Here, $\{\theta,\phi\}$ are respectively spherical polar and azimuthal angles with $\theta = \cos^{-1} (\bk \cdot \uvc{z}/k)$ where $k = |\bk|$ (Fig.~\ref{fig:Fig1}); a modified eigenvalue is defined as $\lambda = (\sigma+\mathrm{i} k \cos\theta + d_T k^2$). Here onwards, we have assumed $\gamma = 1$ corresponding to elongated rod-like particles (see SI). 
The solution to this eigenvalue problem reveals a Hopf bifurcation of the aligned state with a non-trivial dependence on the wavenumber $k$ and the direction of perturbation characterized by the polar angle $\theta$ (Fig.~\ref{fig:Fig1}\textbf{(c)}). In particular, the long-wavelength modes with $k = 0$ are stable (or marginal in the absence of translational diffusion) with the instability emerging at finite $k$. This wavenumber selection is in contrast with the generic alignment instability \cite{ramaswamy2010mechanics} that predicts a scale-free, long-wavelength pitchfork bifurcation of the bulk uniaxial phase \cite{lavi2025nonlinear,lavi2024dynamical}. Furthermore, we find that perturbations that are exactly aligned with the flock ($\theta = 0$) or transverse to the flock ($\theta = \pi/2$) are stable. While the linear theory predicts that for unstable modes, the growth rate increases with $k$ (Fig.~\ref{fig:Fig1}\textbf{(d)}), in practice, such high wavenumbers are cut off due to translational diffusion (or nematic elasticity). Finally, and perhaps most importantly, the predicted instability is absent if $V_s \sim \chi = 0$. Thus, for achiral, spinning particles that do not self-propel, the aligned state is always stable. 

\begin{figure}
    \centering
    \includegraphics[width=0.89\linewidth]{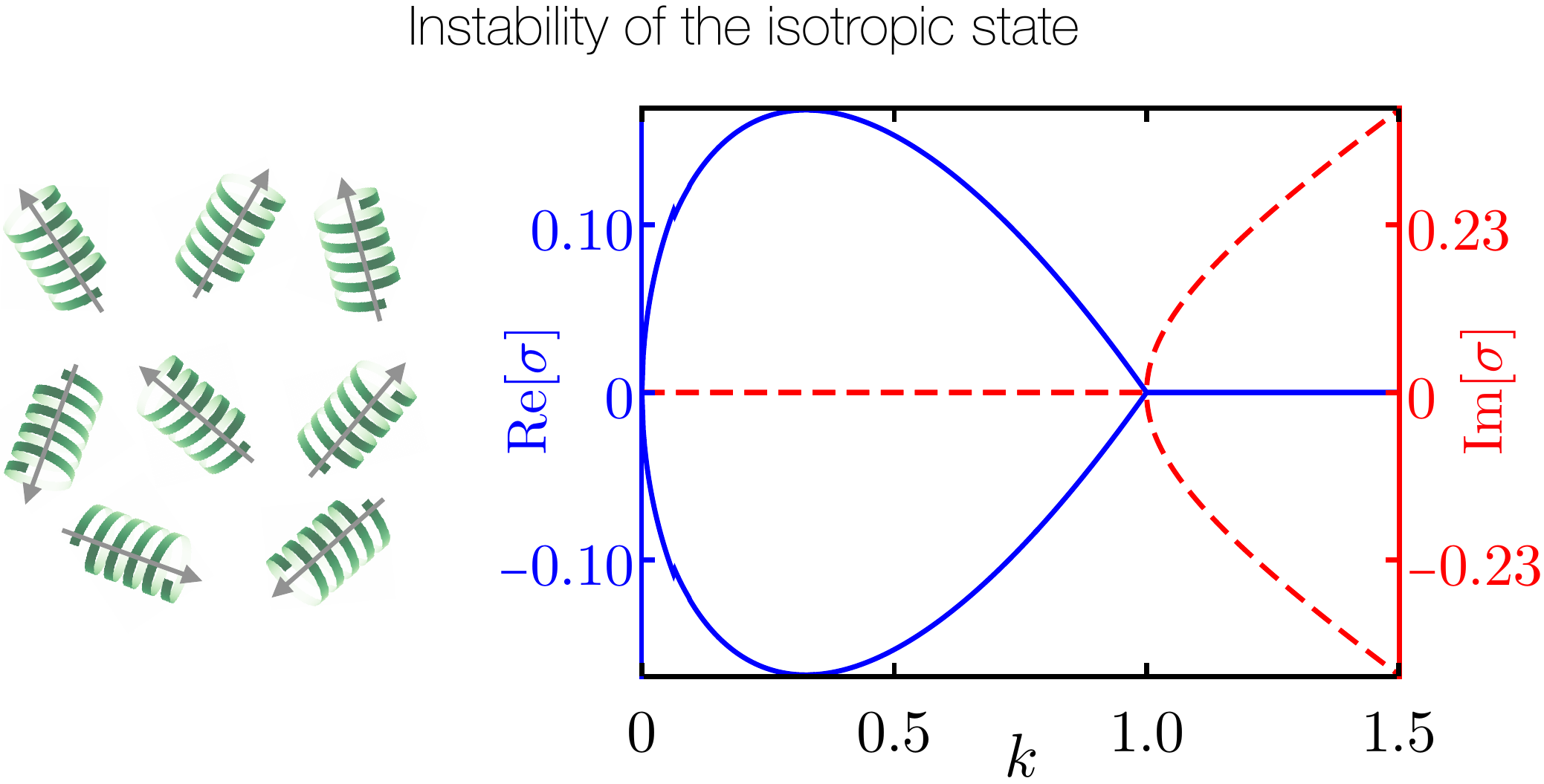}
    \caption{Dispersion relation for the isotropic state depicted pictorially on the left. Here, we have chosen $d_T = d_r = 0$ and $\chi = 1$.}
    \label{fig:Fig2}
\end{figure}

\begin{figure*}
\centering 
\includegraphics[width = 0.95\linewidth]{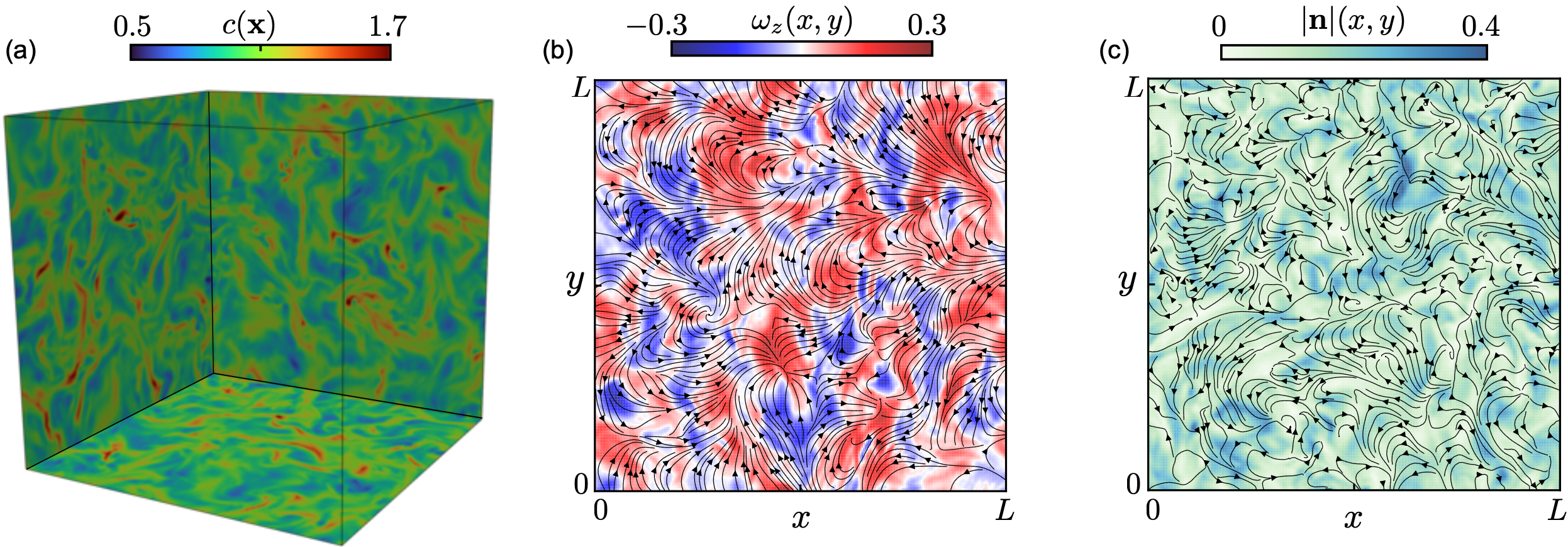}
\caption{Snapshots from a statistically steady state of a 3D direct numerical simulation of the moment equations in a triply periodic square box of size $L$. \textbf{(a)} Concentration field on three faces of the periodic box, revealing fluctuations and formation of bands. \textbf{(b)} Highlights the mean-field velocity streamlines in a 2D slice ($x-y$ plane) from the 3D simulation. The background is color-coded by the out-of-plane vorticity field $\omega_z$. The patches in the vorticity field evolve continuously and exhibit spatio-temporally chaotic patterned states reminiscent of low Reynolds number active turbulence \cite{wensink2012meso}. \textbf{(c)} The polarity field and its magnitude on the same 2D slice as in (b). Starting from an initially isotropic state, the system develops regions of high polarity. \textit{Parameters:} $L = 50 \pi$, $\chi = 0.5$, $d_T=0.01$, $d_r=0.01$; each spatial direction was discretized using $N = 128$ points.}
\label{fig:Fig3}
\end{figure*}

We now probe the stability of another fixed point of our system: the homogeneous, isotropic state of the suspension with $\Psi_0 = 1/4 \pi$ and $\bu(\bx) = \mathbf{0}$. In this isotropic state, we still assume that individual particles spin along their long axis due to external actuations and propel along their backbone $\bp$ as a result of the microscopic chirality. To analyze the stability, we perturb the distribution 
\begin{align}\label{eq:psi_perturb}
    \Psi(\bx,\bp,t) = \frac{1}{4 \pi} \left[1+ \varepsilon \Psi'(\bx,\bp,t) \right], ~~~ |\varepsilon| \ll 1.
\end{align}
Substituting the above expression in Eq.~\eqref{eq:smol_eq} and retaining terms up to $\mathcal{O}(\varepsilon)$ we arrive at the following evolution equation for the perturbation
\begin{align}
    \frac{\partial \Psi'}{\partial t} = - \chi \bp \cdot \nabla_x \Psi' + d_T \Delta \Psi' + 3 \mathbf{pp}:\mathbf{E}',
\end{align}
where $\mathbf{E}'$ is the perturbed rate-of-strain tensor and we have set $d_r = 0$ for simplicity. As above, we consider a plane wave ansatz as $\Psi'(\bx,\bp,t) = \Tilde{\Psi}(\bp,\bk) e^{\mathrm{i}\bk \cdot \bx+\sigma t}$ and obtain a nonlinear dispersion relation (see SI) for unknown growth rate $\sigma$ as
\begin{equation}
    \bigg[2a^2- \frac{4}{3}+(a^3-a) \log \bigg( \frac{a-1}{a+1}\bigg) \bigg] = \pm \frac{4}{3} \chi k,
\end{equation}
where $a=- \mathrm{i}(\sigma +d_T k^2)/ \chi k$. Figure~\ref{fig:Fig2} shows the real and imaginary parts of the numerically obtained growth rate $\sigma(k)$, which—owing to the isotropy of the base state—is independent of wavevector direction. The data reveal a clear wavenumber selection, with the long-wavelength limit ($k=0$) remaining stable. In line with our earlier analysis of the aligned state, the instability emerges only when the particles are self-propelled ($\chi \neq 0$).

\vspace*{1mm}

\noindent {\textit{Features of the instability:}}
Our linear analysis uncovers an alignment instability in both the ordered and isotropic states of the suspension. Two key features distinguish this instability from that in conventional active suspensions: (i) long-wavelength modes are suppressed in the bulk, and (ii) the suspension remains stable in the absence of self-propulsion. The latter feature renders the instability unique to chiral screws, which self-propel as they spin.

To understand the origin of these defining features, we examine the instability of the isotropic state in detail. Velocity fluctuations about this state induce nematic ordering of the particles along the extensional axis of the flow perturbations. The ordering is nematic because Jeffery’s equation, which governs their orientational dynamics, is invariant under the transformation $\bp \to -\bp$ \cite{subramanian2009critical}. In extensile bacterial suspensions, such nematic ordering amplifies velocity fluctuations via dipolar stresses, eventually resulting in an instability. By contrast, in the present system, the particle stress is polar, and velocity fluctuations can only grow if nematic order generates local polar order. Crucially, this nemato–polar coupling arises only for self-propelled particles as previously recognized in \cite{cates2013active}. To illustrate this explicitly, we consider the evolution equations for the concentration $c$, polarity $\bn$, and the nematic order field $\bQ$ given as
\begin{align}
    \partial_t c + \nabla \cdot \left[(\bu + \chi \bn) c\right] - d_T \Delta c &= 0,  \label{eq:con} \\
    (c \mathbf{n})^{\nabla} + \chi \nabla \cdot (c \mathbf{Q}) - d_T \Delta (\mathbf{n}c)  &= - c \mathbf{R}:\mathbf{E}, \label{eq:pol} \\
    (c \mathbf{Q})^{\nabla} + \chi \nabla \cdot (c \mathbf{R}) - d_T \Delta (\mathbf{Q}c) &=-2c\mathbf{S}:\mathbf{E} \label{eq:nem},
 \end{align}
where, $\bQ = \langle \bp \bp \Psi\rangle/c$; $\bR = \langle \bp \bp \bp \Psi \rangle/c$ and $\mathbf{S} = \langle \bp \bp \bp \bp  \Psi \rangle/c$ are higher order moments of the particle distribution and $\langle (.) \rangle = \int_{\bS^2} (.) \md \bp$ denotes orientational averages. We have also defined the upper convected derivative of a vector as $\mathbf{a}^{\nabla} = \mathrm{D}_t \mathbf{a} - \nabla \bu \cdot \mathbf{a} $ and a tensor as $ \mathbf{A}^{\nabla} = \mathrm{D}_t \mathbf{A} - (\nabla \bu \cdot \mathbf{A} + \mathbf{A} \cdot \nabla \bu^T)$. It follows from Eq.~\eqref{eq:pol} that when $V_s \sim \chi = 0$, the polar order remains decoupled from nematic ordering potentially induced by velocity fluctuations. As a result, the isotropic state with $\bn_0 = \mathbf{R}_0 = 0$ remains a stable fixed point of the problem; thus, suspensions of achiral, spinning particles are stable \cite{das2023absence}. Furthermore, at linear order, the growth of polar order is related to fluctuations in $\bQ$ as $\partial_t n_\alpha \sim \chi \nabla_\beta Q_{\alpha \beta} \equiv \mi \chi k_\beta Q_{\alpha \beta}$, where $\bk$ is the wave-vector. This scaling demonstrates that nemato-polar coupling vanishes in the limit $k \to 0$. Thus, long-wavelength velocity fluctuations cannot feed back to polar ordering and, in turn, be amplified, thereby suppressing the scale-free instability typical in dipolar active suspensions \cite{saintillan2008instabilities}.

\vspace*{1mm}

\noindent \textit{Nonlinear simulations:} We next study the long-time dynamics and pattern formation resulting from the instabilities beyond the linear theory. To this end, we perform 3D nonlinear numerical simulations whereby we integrate Eqs.~\eqref{eq:con}-\eqref{eq:nem} in conjunction with the Stokes equation for the mean-field velocity $\bu(\bx)$. The governing equations are solved in a triply periodic domain using a pseudospectral method with a fourth-order accurate Runge-Kutta scheme for time stepping \cite{burns2020dedalus}. Here, we have retained the first three moments of our particle distribution function and have approximated the higher-order moments $\{\bR,\mathbf{S}\}$ using an appropriate closure following  \cite{theillard2019computational,weady2022thermodynamically} (see SI). 

We initiate our simulations close to isotropy with long-wavelength perturbations in all the evolved fields. The box size is chosen such that the unstable wavenumbers are incorporated in the simulation domain. The simulations reveal emergent and self-sustained large-scale flows, emergent polar order, and pronounced concentration fluctuations in the suspension (Fig.~\ref{fig:Fig3}). The collective patterned states with density fluctuations are truly nonlinear and are not predicted by the linear theory. Similar to bacterial suspensions, striation and bands in the concentration field appear because it is advected by the polarity field $\chi \bn$, which is not divergence-free; particles accumulate in regions where $\nabla \cdot (c \bn) < 0$.

\vspace*{1mm}

\noindent \textit{Conclusion:} Through a combination of theory, stability analysis, and large-scale simulations, we have uncovered a previously unknown pathway to pattern formation and collective dynamics in bulk Stokesian suspensions of torque-driven chiral particles. Remarkably, the instability we observe is intrinsically tied to self-propulsion and, therefore, to the microscopic chirality of the constituent particles. This mechanism stands in contrast to well-studied active suspensions, where self-propulsion plays only a minor role \cite{sanchez2012spontaneous} and can even stabilize alignment instabilities \cite{chatterjee2021inertia,ohm2022weakly}. Our micro–macro framework further shows that self-propulsion naturally generates nemato-polar couplings \cite{cates2013active,vafa2025phase}, which are essential for driving this novel class of instabilities. More broadly, we expect our findings will inspire new approaches in the design and exploration of torque-driven active suspensions \cite{chen2025self} with potential implications for transport using actuated nanobots \cite{ghosh2009controlled,chen2025roadmap}. Our theoretical framework enables us to probe the rheology of these suspensions and will be used to develop a mechanistic understanding of the emergent odd viscosity in parity-breaking bulk chiral fluids \cite{fruchart2023odd,2508.04468}.

\vspace*{1mm}

\noindent \textit{Acknowledgments:} 
We thank Anke Lindner and Jasmin Imran Alsous for feedback on this work. B.C. acknowledges the support of the Department of Atomic Energy, Government of India, under project no. RTI4001. Part of the work was carried out at ESPCI, Paris, supported by the SSHN fellowship (2024). S.C. thanks Vinay Kumar and Santhiya P S for their help with Dedalus and Paraview. 

\bibliography{bibfile}
\end{document}